\documentstyle[12pt]{article}
\advance\textheight by \topskip
\begin{document}
\begin{flushright}
CERN-TH/95-211\\
SU-ITP-95-16\\
hep-th/9508072\\
August 15, 1995\\
\end{flushright}
\vspace{1cm}
\begin{center}
\baselineskip=16pt

{\Large\bf  ${\bf N=2}$\, EXTREMAL BLACK HOLES}  \\

\vskip 2cm
 {\bf Sergio Ferrara}\footnote{E-mail:
ferraras@cernvm.cern.ch}\\
 \vskip 0.2cm
Theory Division, CERN, 1211 Geneva 23, Switzerland\\
\vskip .6cm

{\bf Renata Kallosh}\footnote {E-mail:
kallosh@physics.stanford.edu}\\
 \vskip 0.2cm
Physics Department, Stanford University, Stanford   CA 94305-4060, USA\\
\vskip .6cm

 {\bf Andrew Strominger}\footnote{E-mail:
andy@denali.physics.ucsb.edu}\\
 \vskip 0.2cm
Department of Physics,    University of California, Santa Barbara,   CA
93206-9530, USA\\

\vskip 1 cm

\end{center}
\vskip 1 cm
\centerline{\bf ABSTRACT}
\begin{quotation}

It is shown that extremal magnetic black hole solutions of N = 2
supergravity coupled to vector multiplets $X^\Lambda$ with a generic
holomorphic prepotential $F(X^\Lambda)$ can be described as
supersymmetric solitons which interpolate between maximally symmetric
limiting solutions at spatial infinity and the horizon. A simple
exact solution is found for
the special case that the ratios
of the $X^\Lambda$ are real, and it is seen that the logarithm of
the
conformal factor of the spatial metric equals the K\"{a}hler potential
on the vector multiplet moduli space. Several examples are
discussed in detail.
\end{quotation}
\newpage
\baselineskip=15pt
\section{Introduction}

Black holes seem to be a never-ending source of surprises.
While much has been learned about their behavior, much remains to
be understood  --  even at the classical level. In this paper we study
the classical supersymmetric solutions in a general theory with
N = 2 supersymmetry. Previous work on this subject can be found in
 \cite{GH,T}. The solutions appear to have a richer structure
than the more thoroughly studied N = 4 case. In section 2 we recall some
aspects of N = 2 supergravity. In section 3 the magnetic  solutions are
described in terms of trajectories in the  special geometry of the N = 2
moduli  space which terminate at a supersymmetric fixed point at the
horizon.  In section 4 we find that the equations can be
integrated for  a restricted but large class of cases.  An intriguing
relation between  the K\"{a}hler potential on the moduli space and the metric
conformal factor emerges. Some simple examples are worked out in detail in
section 5.  We do not attain a complete characterization  of the
classical geometry of N = 2 black holes in this paper, but we hope
that our results prove useful for future efforts in this direction.

\section{Special Geometry and N = 2 Supersymmetry}
We study N = 2 supergravity coupled to $n$ \ N = 2 vector multiplets in the
framework
of special geometry \cite{CREM}--\cite{CAFP}. In this section
some formulae that will be needed in the following are recalled.
Further details can be found in \cite{WLP} whose notation we adopt.
The supergravity theory is defined in terms of a projective  covariantly
holomorphic section $(X^\Lambda(\phi^i), -{i\over 2}F_\Lambda(\phi^i))$,
$\Lambda = 0, 1, ..., n, ~~i=1,...,n$, of an ~$Sp(2n+2)$ vector bundle over the
moduli space parametrized by $\phi^i$. (We note that alternate conventions
are often employed in which the  definition of $F_\Lambda$ differs by a factor
of $2i$.)
In some cases the theory can be described in terms of a covariantly
holomorphic function
$F(X)$
of degree two:
\begin{equation}\label{2}
F_\Lambda(\phi^i) = F_\Lambda(X(\phi^i)) = {\partial\over
\partial X^\Lambda} F(X) \ .
\end{equation}
 Given a prepotential $F(X)$, or a covariantly holomorphic section
$(X^\Lambda,\, -{i \over 2} F_\Lambda)$, one can construct the entire scalar
and vector  parts of the action.

It is convenient to introduce the
inhomogeneous coordinates
\begin{equation}\label{10}
Z^\Lambda = {X^\Lambda(\phi_i)\over X^0(\phi_i)} \ , ~~~~~~
Z^0 = 1 \ .
\end{equation}
We assume $Z^\Lambda(\phi_i)$ to be invertible, so that, in special
coordinates, ${\partial Z^\Lambda\over
\partial
\phi^i}  =\delta^\Lambda_i$. ~In this case the complex scalars $Z^i = \phi^i$
{}~($
i =
1,...,n$) represent the lowest component of the $n$ vector multiplets of N = 2
supersymmetry. The K\"{a}hler potential determining the metric of these fields
is
\begin{eqnarray}\label{22}
&&K(Z,\bar Z)=2\ln |X^0| = -\ln \left (N_{\Lambda\Sigma}(Z,\bar Z)\,
Z^\Lambda\bar Z^\Sigma\right )\nonumber \\
&&= - \ln {1\over 2} [f(Z) + \bar f(\bar Z) + {1\over 2} (Z^i - \bar Z^i) (\bar
f_i - f_i)]
\ ,
\end{eqnarray}
where $N_{\Lambda\Sigma} = {1\over  4}(F_{\Lambda\Sigma} + \bar
F_{\Lambda\Sigma})$ and $f(Z) = (X^0)^{-2} F(X)$.
In the conformal gauge \cite{WLP}
\begin{equation}\label{conf}
N_{\Lambda\Sigma} X^\Lambda \bar X^{\Sigma}=1 \ .
\end{equation}
The graviphoton field strength, as well as the field strengths of the $n$
Abelian vector
multiplets, are constructed out of $n+1$ field strengths $\hat
F^\Lambda_{\mu\nu}
= \partial_\mu W^\Lambda_\nu - \partial_\nu W^\Lambda_\mu$. The graviphoton
field  strength is
\begin{equation}\label{23}
T_{\mu\nu}^{+}  = {4 N_{\Lambda\Sigma} X^\Lambda\over N_{IJ} X^I X^J} \  \hat
F_{\mu\nu}^{+\Sigma} \ ,
\end{equation}
where the superscript + (-) denotes the (anti)-self-dual part. This
defines the central charge of the theory, since it enters into
gravitino
transformation rules. The vector field strengths which enter the gaugino
supersymmetry
transformations
are
\begin{equation}\label{24}
{\cal F}_{\mu\nu}^{+\Lambda} = {\hat F}_{\mu\nu}^{+\Lambda}
-{1\over 4} X^\Lambda\, T_{\mu\nu }^{+} \ .
\end{equation}
The anti-self-dual vector field strengths ${\hat
F}^{-\Lambda}$ are part of the symplectic vector
\begin{equation}\label{11}
 \pmatrix{
{\hat F}^{-\Lambda} \cr -2i
\bar { \cal N}_{\Lambda\Sigma}\,{\hat F}^{-\Sigma} \equiv
i G^-_{\Lambda}
\cr }  ,
\end{equation}
where again the expression for the  matrix
${\cal N}_{\Lambda\Sigma}$ is
derived from the prepotential $F(X)$ \cite{WLP}--\cite{CAFP}. The vector part
of the
action is then
proportional to
\begin{equation}\label{12}
\mbox{ Re}\ {\hat F}^{-\Lambda} G^-_\Lambda \ ,
\end{equation}
and the graviphoton field strength can be written in the manifestly symplectic
form
\begin{equation}\label{tst}
T_{\mu\nu}^-  =  2X^\Lambda G^-_{\Lambda\mu\nu}+ F_{\Lambda}
{\hat F}_{\mu\nu}^{-\Lambda} \ .
\end{equation}
The Lagrangian for the scalar components of the vector multiplets is
defined by
the K\"{a}hler potential as
\begin{equation}\label{13}
g_{i\bar j}\,\partial_\mu\phi^i\, \partial_\nu\bar\phi^{\bar j}\, g^{\mu\nu} \
,
\end{equation}
where $g^{\mu\nu}$ is the space-time metric, and
\begin{equation}\label{14}
g_{i\bar j}= \partial_i \partial_{\bar j} K(\phi, \bar\phi) \ .
\end{equation}

The gravitino supersymmetry transformation law, to leading order in fermi
fields, is
\begin{equation}\label{grt}\delta \psi^\alpha _\mu=2\nabla_\mu\epsilon^\alpha
-{1 \over 16}\gamma^{\nu\lambda}T^-_{\nu\lambda}\gamma_\mu
\epsilon^{\alpha \beta}
\epsilon_\beta+iA_\mu\epsilon^\alpha \ ,
\end{equation}
where $ \alpha, \beta=1,2$ are $SU(2)$ indices and  $A_\mu= {i \over 2}
N_{\Lambda\Sigma} [\bar X^\Lambda \partial_\mu  X^\Sigma
- (\partial_\mu  \bar X^\Lambda) X^\Sigma]$.
The gaugino transformation law is
\begin{equation}\label{ggt}\delta \Omega^\Lambda_\alpha
=2\gamma^\mu\nabla_\mu X^\Lambda
 \epsilon_\alpha
+{1 \over 2}\gamma^{\nu\lambda}
{\cal F}^{+\Lambda}_{\nu\lambda} \epsilon_{\alpha \beta}
\epsilon^\beta+2i\gamma^\mu A_\mu\epsilon_\alpha \ .
 \end{equation}

BPS states of the N = 2 theory have a mass equal to the central charge $z$.
It follows from the supersymmetry transformation rules that this is simply
the graviphoton charge
\cite{CAFP}
\begin{equation}\label{15}
M = |z|  =
  |q^{(e)}_\Lambda X^\Lambda - {i \over 2} q_{(m)}^\Lambda F_\Lambda|=
e^{K/2} |q^{(e)}_0 +  q^{(e)}_i Z^i + {i \over 2} (q_{(m)}^0 Z^i -
q_{(m)}^i)f_i - iq^0_{(m)} f|
\ ,
\end{equation}
where $q^{(e)}$ and $q_{(m)}$ are electric and magnetic charges
associated to $i\cal G$ and $\hat F$ and comprise a symplectic vector.
Duality transformations of the N = 2 theory  correspond to
different choices of the symplectic representative
$(X^\Lambda,-{i \over 2}F_\Lambda)$ of
the symplectic geometry.

\section{ Magnetic N = 2 BPS Black Holes}
In this section we discuss the general form of the supersymmetric magnetic
black hole solutions and their interpretation as interpolating solitons.
It has been shown by Tod \cite{T} in general that, for N = 2 theories,
a static metric admitting supersymmetries can be put in the form
\begin{equation}\label{17}
ds^2 = -e^{2U} dt^2 + e^{-2U} d\vec x^2 \ .
\end{equation}
For spherically symmetric  black hole solutions $U$ will be a function
only of the radial coordinate $r$. By solving the Bianchi
identities $d \hat F^\Lambda = 0$ we then find for the
radial component of the magnetic\footnote{ It follows from equation (7) that
for a
generic prepotential $F$ the electric
charge will be nonzero despite the fact that  $\hat
F$ is a magnetic field strength.
In general this electric charge will not respect the charge
quantization condition.
This can be avoided by restricting the prepotential as in the examples in
section 5.
} field strengths $ \hat F^\Lambda_r
\equiv 2{\epsilon_r}^{\theta \phi} \hat F^\Lambda_{\theta \phi}$ :
\begin{equation}\label{26}
\hat F^\Lambda_r = {q^\Lambda\over r^2} \, e^{U(r)} \ .
\end{equation}
Inserting (\ref{17}) and (\ref{26}) into the gravitino transformation law,
and demanding that the variation vanish for some choice of $\epsilon$,
we derive the following first order
differential equation:
\begin{equation}\label{27}
4 U'= -\sqrt{(\bar ZNq) (ZNq) (\bar Z N Z)\over (ZNZ)
(\bar Z N \bar Z)}~e^{U}\ ,
\end{equation}
where $U' \equiv {\partial U\over \partial \rho}$, $\rho\equiv 1/r$,
and we employ the notation $(ZNq)\equiv Z^\Lambda N_{\Lambda\Sigma}\,q^\Sigma$.

Equation (\ref{27})
may be viewed as determining
$U$ as a function of the moduli fields $Z^\Lambda$.
A vanishing \ gaugino
transformation further requires that the moduli fields obey
\begin{equation}\label{28}
(Z^\Lambda)'   = -{e^{U}\over 4}\, \sqrt{(ZNZ) (\bar Z Nq) (\bar Z NZ)
\over (\bar Z N \bar Z) (ZNq)}~
\Bigl(
 Z^\Lambda q^0  - q^\Lambda\Bigr) \ .
\end{equation}
Differentiating again with respect to $\rho$ and substituting (\ref{27})
leads to the second order differential equation
\begin{equation}\label{28g}
(Z^\Lambda)''-\left(
{ (ZNq)   \over
 (ZNZ)  }+q^0 \right)
{((Z^\Lambda )')^2 \over  Z^\Lambda q^0  - q^\Lambda}
+{1 \over 2} \left(\ln {(ZNZ) (\bar Z Nq) (\bar Z NZ)
\over (\bar Z N \bar Z) (ZNq)}
\right)' (Z^\Lambda )'=0\ .
\end{equation}
This equation is independent of $U$ and
can be viewed as a generalized geodesic equation which describes how $Z$
evolves as one moves into the core of the black hole. Initial conditions
for $Z$ are  specified at infinity ($\rho=0$) corresponding to the
asymptotic values
of the field. The first derivative of $Z$ is then fixed in terms of  the
charge of the black hole by the supersymmetry constraint (\ref{28}). $Z$
will then evolve until it runs into a fixed point. It is evident from
(\ref{28}) and (\ref{28g}) that these fixed points are at
\begin{equation}\label{zf}
Z^\Lambda_{fixed}={q^\Lambda \over q^0} \ ,
\end{equation}
where $(Z^\Lambda)'=0$. Each fixed point  is typically
surrounded by a
finite basin of attraction. The limiting form of the metric  at
such a fixed point is found by integrating (\ref{27}),
\begin{equation}\label{uc}
e^{-U}={c \over 4}\, \rho \ ,
\end{equation}
where the constant $c$ is given by
\begin{equation}\label{cv}
c ={\sqrt{ q^\Lambda N_{\Lambda\Sigma}\,q^\Sigma }}
=q^0\, e^{-K(Z^\Lambda_{fixed} )/2}\ .
\end{equation}
This corresponds to the maximally symmetric charged Robinson-Bertotti universe.
Thus, as in  \cite{GT}, the extremal black holes may be viewed as
solitons which interpolate between maximally symmetric vacua at infinity and
the horizon.

The locations of the fixed points (\ref{zf}) depend on the charges
but not on the asymptotic values of the moduli fields. Thus if the
the asymptotic values of those fields are adiabatically changed, the
geometry of the black hole near the horizon remains fixed.
Symplectic invariance implies a similar structure for many of the dyonic and
electrically charged extremal black holes.

\section{Space-Time Geometry  from K\"{a}hler  Geometry}

In this section we consider a remarkably simple special class of solutions
which exist for a generic prepotential. We will work in a symplectic basis in
which $q^0=0$
in order to simplify the equations.
In such a basis the fixed points $Z^\Lambda_{fixed}$
move to infinite coordinate values. Thus solutions for which the moduli field
is constant at the fixed point (which corresponds to the Reissner-Nordstr\"om
solution) cannot be described in this basis\footnote{However, by  using the
manifestly symplectic constraint  \cite{CAFP} instead of the superconformal one
 \cite{WLP}, one can describe the Reissner-Nordstr\"om configuration  in terms
of the K\"{a}hler potential, see example 5.4.2}.
In the $q^0=0$ basis it is straightforward to check that (\ref{27}) and
(\ref{28}) are solved by
\begin{equation}\label{31}
e^{2U(\rho)}=
e^{K(Z,\bar Z)- K_\infty} \ ,
\end{equation}
and \begin{equation}\label{harmreal}
 Z^i  = Z^i_{\infty} + {q^i \over 4} \rho \, e^{-K_\infty/2} \ ,
\end{equation}
provided that the asymptotic value of $Z^i$ is restricted to obey
\begin{equation}\label{hrm}
 Z^i_{\infty}  = \bar Z^i_{\infty}  \ .
\end{equation}
Alternatively we may have
\begin{equation}\label{harmimagin}
 Z^i  = Z^i_{\infty} + i {q^i \over 4} \rho \, e^{-K_\infty/2} \ ,
\end{equation}
provided that the asymptotic value of $Z^i$ is restricted to obey
\begin{equation}\label{hrm2}
 Z^i_{\infty}  = - \bar Z^i_{\infty}  \ .
\end{equation}
The restrictions imply that these solutions exist only at special points
in the moduli space where all $Z^i$ are either real or imaginary. More general
solutions
may be obtained from these by symplectic transformations.

Thus the space-time metric is
\begin{equation}\label{34}
ds^2 = e^{K(Z^i,\bar Z^i=Z^i) - K_\infty}\, dt^2 -
e^{-K(Z^i,\bar Z^i=Z^i) + K_\infty}\,
d\vec x^2 \ ,
\end{equation}
where each $Z^i$ solves a  three-dimensional  harmonic equation and is given in
eq. (\ref{harmreal}) or  (\ref{harmimagin}).
Hence the logarithm of the spatial conformal factor is identified with the
moduli space K\"{a}hler potential.

\section{Examples of\, ${\bf N \geq 2}$\, BPS states}
\subsection{ Calabi-Yau  magnetic black holes}
The prepotential is
\begin{equation}\label{44}
F = i d_{ABC}\, {X^A X^B X^C \over X^0} \ .
\end{equation}
We consider pure imaginary $Z^A$ and real $d_{ABC}$ (corresponding to the
classical Calabi-Yau moduli space)
\begin{equation}
\qquad   e^{- K(Z,\bar Z)} =  - 2  d_{ABC}  \,  {\rm Im} Z^A \, {\rm Im} Z^B \,
{\rm Im} Z^C,
\end{equation}
where ${\rm Im} Z^A  = {1\over 2i} (Z^A - \bar Z^A)={\rm Im} (Z^A)_\infty + ~
{q^A_{(m)}\over
 r}\, e^{-K_\infty/2}$.
\begin{equation}\label{45}
ds^2 = \left(  { \, d_{ABC}  \,  {\rm Im} Z^A \, {\rm Im} Z^B \, {\rm Im} Z^C
\over  \, [d_{ABC}  \,  {\rm Im} Z^A \, {\rm Im} Z^B \, {\rm Im} Z^C]_\infty
}\right )^{-1}\, dt^2 -
\left(  { \, d_{ABC}  \,  {\rm Im} Z^A \, {\rm Im} Z^B \, {\rm Im} Z^C \over
\, [d_{ABC}  \,  {\rm Im} Z^A \, {\rm Im} Z^B \, {\rm Im} Z^C]_\infty  }\right
)\,
d\vec x^2 \ .
\end{equation}

\subsection{Massive and massless  ${\bf SU(1,n) \over SU(n)} $  supersymmetric
white holes}
The prepotential is
\begin{equation}\label{36}
F(X^0,X^1) = (X^0)^2- (X^i)^2 \ , \qquad  e^{- K(Z,\bar Z)} = 1-|Z^i|^2 \ .
\end{equation}
Here  the $Z^i$ are real:
\begin{equation}\label{38}
 Z^i   = Z_\infty^i  + {q^i \over r}\, e^{-K_\infty/2} \ ,
\end{equation}

\begin{equation}\label{39}
ds^2 = \left({1-|Z|^2\over 1-|Z_\infty|^2}\right)^{-1}\, dt^2 -
\left({1-|Z|^2\over 1-|Z_\infty|^2}\right)\,
d\vec x^2 \ .
\end{equation}
In particular  in the simplest case of $i=1$ we get
\begin{equation}\label{40}
g_{tt} = g_{rr}^{-1} =  \left(1- {2 Z_\infty\, q\over r}\,e^{K_\infty/2} -
{q^2\over
r^2}\right)^{-1} .
\end{equation}
To satisfy the supersymmetry bound we require
\begin{equation}\label{41}
 M =-  {Z_\infty \, q\over\sqrt{1-|Z_\infty|^2}} \geq 0 \ ,
\end{equation}
and the geometry is
\begin{equation}\label{42}
g_{tt} = g_{ii}^{-1} = \left(1+{2 M\over r}  - {q^2\over
r^2}\right)^{-1}.
\end{equation}
These configurations are non-trivial in the limit when the ADM mass tends to
zero.
It will be interesting to understand if or when such states can arise in a
physical
theory.
The limit $M \to 0$ can be achieved via $Z_\infty \to 0$. In this limit the
scalar field
becomes inversely proportional to the radius $r$,
\begin{equation}\label{43}
Z(r) = {q\over  r} \ .
\end{equation}

For N = 4 BPS states the analogous massless states have been
studied recently \cite{Klaus2}--\cite{CYHETS}.  The configuration exhibits a
repulsive (i.e. antigravitating) singularity and was referred to as a
supersymmetric white hole \cite{KL}.
The difference between the metric (\ref{42}) for N= 2 and the corresponding
metric obtained in \cite{Klaus2}--\cite{KL}\,  for N = 4 is that $g^{N =
4}_{tt} =
(g^{N = 2}_{tt})^{1/2}$.
This does not change the repulsive  nature of the singularity.

 \subsection{$\bf { SO (2,1)\over SO(2) } \times  {SO(2,n) \over SO(2) \times
SO(n)}$\,  BPS states}
Here again we consider pure imaginary $Z$.
\begin{equation}\label{48}
F = -i\, {X^s \over X^0}\, \Bigl((X^1)^2 - \sum^n_{a = 2}(X^a)^2 \Bigr) \ ,
\qquad
e^{- K(Z,\bar Z)} = 2 {\rm Im} Z^s\, \Bigl(( {\rm Im} Z^1)^2 - \sum^n_{a= 2}(
{\rm Im} Z^a)^2 \Bigr)\ .
\end{equation}

The metric of the BPS configuration  defined by the K\"{a}hler potential is
given by
\begin{equation}\label{49}
g_{tt} = g^{-1}_{ i  i} = \left[{ {\rm Im} Z^s \Bigl(({\rm Im} Z^1)^2 -
({\rm Im} Z^a)^2\Bigr)\over
\left[ {\rm Im}  Z_s \Bigl(({\rm Im} Z^1)^2 - ({\rm Im}
Z^a)^2\Bigr)\right]_{\infty} }\right]^{-1}.
\end{equation}

The configurations presented in this example may contain both types of
supersymmetric states, those with attractive singularities and those with the
repulsive ones, depending on the choice of the parameters describing the
harmonic functions
\begin{equation}\label{46a}
  Z^i  = Z^i_\infty + ~ i {q^i_{(m)}\over
 r}\,e^{-K_\infty/2}\ ,~~~~~ i = \{ s, 1, a=2, \dots , n \}   .
\end{equation}
 Note that this example actually provides one of the particular choices of
Calabi-Yau  magnetic black holes.

\subsection{$\mbox{N=4, 2}$  pure supergravity black  holes from  K\"{a}hler
geometry perspective}
Here we will analyse some previously known\, supersymmetric black hole
solutions in the framework of the manifestly symplectic formalism \cite{CAFP}.

The K\"{a}hler potential  is different from the conformal gauge (\ref{conf})
(here $X_\Lambda, \, F_\Lambda $ are holomorphic
sections) and is
given by
\begin{equation}\label{a7}
e^{- K(X, \bar X) } =  \bar X^\Lambda N_{\Lambda \Sigma}  X^\Sigma \ .
\end{equation}
\subsubsection{$\mbox {SL(2,Z)}$ axion-dilaton dyons}
 It is instructive to analyse the known
  $SL(2,Z)$-invariant axion-dilaton dyonic
black hole solution  \cite{KO} of N = 4 supergravity as a particular solution
of N = 2
supergravity coupled to an N = 2  vector multiplet.
These solutions have complex moduli. The prepotential for this
theory is $F(X) = 2  X^0 X^1$.
The holomorphic section includes
\begin{equation}\label{a4}
\left (\matrix{
X^\Lambda\cr
-{i\over 2} F_\Lambda \cr
}\right )  \Longrightarrow \left (\matrix{
X^0\cr
-iX^1\cr
}\right ),  ~~ \left (\matrix{
X^1\cr
-iX^0
\cr
}\right ).
\end{equation}
For the axion-dilaton black hole $X^0$ and $X^1$ can be  identified with two
complex harmonic functions ${\cal H}_1$, ${\cal H}_2$ as follows:
\begin{equation}\label{a6}
X^1 =  {\cal H}_1(\vec x) \ , ~~~~~ X^0 =  i {\cal H}_2(\vec x) \ .
\end{equation}
The K\"{a}hler potential is
\begin{equation}\label{a7a}
e^{- K(X, \bar X) } = i \,(\bar {\cal H}_1 {\cal H}_2  -\bar {\cal H}_2
{\cal H}_1)  = g_{tt}^{-1}(\vec x) \ ,
\end{equation}
in agreement with the metric of the axion-dilaton black hole found
in \cite{KO}.
\subsubsection{Reissner-Nordstr\"om solution}
 The prepotential for  pure $N=2$ supergravity  is $F(X) = ( X^0 )^2$.
The K\"{a}hler potential of this theory in the manifestly symplectic formalism
\cite{CAFP} is given by
\begin{equation}\label{a7aa}
e^{- K(X, \bar X) } = X^0 \bar X^0 = V^{-1} (\vec x ) \bar V^{-1} (\vec x )
=e^{-2U (\vec x)}\ ,
\end{equation}
where $X^0=V^{-1} (\vec x )$ is a real  (imaginary) harmonic function for
electric (magnetic) Reissner-Nordstr\"om extremal supersymmetric black hole
\cite{GH,T}.

\vskip .5 cm

In conclusion, we have found a   simple  relation between the special geometry
describing the couplings of scalars and vectors in extended
 locally supersymmetric theories and the space-time geometry of the
black-hole-type solutions  in these theories. It is likely that more general
solutions with complex moduli
will be found in this framework.

\section*{Acknowledgements}
We are grateful to the Aspen Center for Physics where this work was
initiated.  S.F. was supported by DOE
 grants DE-AC0381-ER50050 and DOE-AT03-88ER40384,Task E, and by EEC
 Science Program SC1*CI92-0789.\, R.K. was  supported
by  NSF grant PHY-8612280.\, A.S. was supported by DOE grant DOE-91ER40618.


\end{document}